\documentclass[twocolumn]{aastex701}
\usepackage{graphicx}
\usepackage{amsmath}
\usepackage{subcaption}
\usepackage{tabularx}
\usepackage{float}
\usepackage{tikz}
\usepackage[utf8]{inputenc}
\usepackage{textgreek}
\usepackage[percent]{overpic}
\usetikzlibrary{arrows.meta,positioning,shapes.misc,fit,shapes}

\tikzset{
 stage/.style={
 rectangle, rounded corners=2mm, draw=black, fill=gray!6,
 very thick, text width=0.9\columnwidth, align=left, inner sep=6pt
},
 outbox/.style={
 ellipse, draw=black, fill=white, very thick,
 align=center, inner sep=2.5pt
 },
 connector/.style={-Stealth, very thick},
 badge/.style={
 circle, draw=black, fill=black, minimum size=6mm,
 inner sep=0pt, font=\bfseries, text=white
 }
}

\begin{document}

\title{Mapping Oscillatory Flows in a Giant Chromospheric Spiral}

\author[0009-0009-2284-462X]{Yash.B.Saneshwar}
\email{yash.saneshwar@northumbria.ac.uk}
\affiliation{Northumbria University, \\
Newcastle upon Tyne, NE1 8ST, UK}

\author[0000-0001-9590-6427]{Eamon Scullion}
\email{eamon.scullion@northumbria.ac.uk}
\affiliation{Northumbria University, \\
Newcastle upon Tyne, NE1 8ST, UK}

\author[0000-0002-5915-697X]{Gert J. J. Botha}
\email{gert.botha@northumbria.ac.uk}
\affiliation{Northumbria University, \\
Newcastle upon Tyne, NE1 8ST, UK}

\begin{abstract}
The solar chromosphere is permeated by complex magnetic fields that guide plasma flows and energy into the corona. This work presents a detailed analysis of a unique, high-resolution observation of a giant chromospheric spiral structure that emerges due to a large magnetic pore, captured by the Swedish 1-m Solar Telescope (SST). A comprehensive data analysis pipeline is developed to automatically detect the edges of 2255 plasma flows (loops) that constitute the spiral, and these are used to extract the kinematics of flows propagating along the magnetic field. The analysis reveals three primary insights into the spiral's physics. First, magnetic curvature is correlated with oscillatory flow dynamics, i.e. regions of high loop curvature exhibit a statistically significant excess of higher-order oscillation modes compared to straighter loops; it is also correlated with higher intensity and longer periods. Second, spatial distribution of oscillation period shows an inverse trend, decreasing from $\sim$3.5 minutes in the pore to $\sim$3 minutes in the outer spiral arms. This is interpreted as a signature of the overlying trans-equatorial quadrupolar coronal loop system compressing the pore's field lines into a near-horizontal orientation, producing a period gradient that challenges the standard expanding canopy model. Finally, the emission signature confirms that oscillating threads represent localised channels of brightness that lie within cooler, absorbing loop material. This study provides the first statistical analysis of oscillatory flows in a large-scale spiral, probing energy flow through the chromosphere through curved magnetic structure.

\end{abstract}


\section{Introduction} \label{sec:intro}

The transport of energy from the Sun's convection zone into the chromosphere and corona remains a central problem in solar physics. The magnetic field acts as a waveguide along which MHD disturbances propagate \citep{Klimchuk_2006}. Since the identification of Alfv\'enic waves \citep{Alfven1942_Nature,DePontieu2007}, a range of mechanisms have been proposed that dissipate wave energy, including resonant absorption and phase mixing \citep{Ionson1978_ResAbs,Heyvaerts_1983}. Observations and reviews over the last two decades show that Alfv\'enic and magnetoacoustic waves are widespread throughout the chromosphere and corona \citep{Tomczyk2007_Science,DeMoortelNakariakov2012_Review,Mathioudakis2013_SSR,Jess2015_SSR,KhomenkoCollados2015_LRSP}.

In the chromosphere, fibrils and narrow flow channels that can be seen in strong chromospheric lines are often used as proxies for the projected magnetic-field geometry, although their correspondence with field lines is not always one to one \citep{delaCruzRodriguez2011,Leenaarts2015}. Therefore the measured wave properties can be used to infer properties of the magnetic structures that are the guiding medium, when direct magnetic diagnostics are limited \citep{DeMoortelNakariakov2012_Review,Jess2015_SSR,KhomenkoCollados2015_LRSP}. In a standard sunspot model it is seen that the dominant wave period varies systematically with magnetic inclination and atmospheric stratification, with more inclined fields supporting longer-period oscillations \citep{Jefferies2006,Jess2013,Jess2015_SSR,KhomenkoCollados2015_LRSP}. This behaviour is often discussed in terms of the inclination-dependent modification of the effective acoustic cut-off frequency, which allows longer-period waves to propagate along inclined fields \citep{BelLeroy1977,McIntoshJefferies2006,Yuan2014, Scullion_2009}.
These relationships suggest that spatial variations in wave properties can act as proxies for changes in field geometry and the local propagation environment.

Wave mode conversion provides one pathway by which wave energy can be redistributed, particularly near the equipartition layer where the sound and Alfv\'en speeds are comparable \citep{Miriyala_2025}. Foot-point driven slow magnetoacoustic waves can undergo mode conversion as they encounter regions of varying plasma-$\beta$, with the efficiency depending on the angle between the wave vector and the magnetic field \citep{Cally2008}.

High-resolution observations also show that chromospheric fine structure can organise into rotating and spiral-like patterns. Quiet-Sun swirl events and ``magnetic tornadoes'' have been interpreted as signatures of twisted or rotating magnetic structures that can channel energy upward, and they often show clear intensity and flow organisation in strong chromospheric lines \citep{Wedemeyer-Bohm_2009, Wedemeyer-Bohm2012, Shetye_2019, Tziotziou_2023}. These structures provide a natural environment for geometry-driven oscillatory behaviour, because the guiding field can vary rapidly in curvature and inclination over relatively small distances. On larger scales, giant solar tornadoes have been observed to extend from the chromosphere into the corona, representing a macro-scale analogue of such rotating magnetic structures and further highlighting the role of twisted field geometries in energy transport through the solar atmosphere \citep{Li_2012}. 

Small-scale chromospheric swirls are now known to be ubiquitous features of the quiet solar atmosphere. Statistical studies using automated detection applied to high-resolution H$\alpha$ SST/CRISP observations find a mean surface density of $\sim$0.08 swirls~Mm$^{-2}$ and an occurrence rate of $\sim$10$^{-2}$ swirls~Mm$^{-2}$~min$^{-1}$, with radii of 0.5--2.5~Mm and mean lifetimes of $\sim$10~min \citep{Dakanalis2022}. These swirls, however, are the product of a different mechanism from the structure studied here: they result from the interaction of small-scale magnetic flux concentrations with intergranular vortex flows, whereas the giant $\sim$20-Mm spiral reported here is anchored to a magnetic pore and connected to a much larger flux/loop system. Its formation is instead tied to the rotation of the host pore, driven by the emergence of twisted sub-photospheric flux tubes, which is a commonly observed phenomenon in active regions \citep{Brown2003, Sturrock2015}. Large-scale spiral structures of this spot- and pore-related kind are themselves not unprecedented in the chromosphere: the spiral topology of H-$\alpha$ fibrils and filaments around sunspots has long been recognised and shown to be consistent with a force-free magnetic field geometry \citep{Nakagawa1971}, and \citet{Schmieder1989} found that a single value of the force-free parameter $\alpha$ reproduces both a sheared filament structure and the surrounding sunspot spiral structures, indicating that both are consistent with the same linear force-free magnetic field model. As a distinct but related oscillatory phenomenon, one- and two-armed spiral wavefront patterns interpreted as slow magnetoacoustic waves guided upward along inclined magnetic fields have also been observed within sunspot and pore umbrae \citep{Su2016, Felipe2019, Kang2019}. Taken together, these studies show that both large-scale spiral chromospheric morphologies and spiral wavefront patterns have been observed repeatedly in association with sunspots and pores. We therefore interpret the present structure as observationally rare, considerably larger and longer-lived than the common swirl population and captured here at high resolution over its full 37-minute lifetime, rather than unique. The recurrence of similar large-scale spiral chromospheric morphologies indicates that this class of magnetic topology is not unprecedented, although the frequency with which such structures occur remains unknown. A statistical census of such large-scale structures has yet to be carried out, making the present study a vital first step toward characterising their oscillatory dynamics.
 
Specifically, we focus on an observation of a long-lived giant spiral structure in the chromosphere anchored by a magnetic pore. Pores are concentrations of strong magnetic flux and represent a fundamental stage in sunspot formation \citep{Sobotka_1999}. The spiral pattern is formed by fine-scale channels along which plasma flows outward from the pore and within which transverse oscillatory motions are observed (specifically, the propagation of longitudinal oscillatory flows along the loops, observed as intensity variations in space--time diagrams, rather than transverse spatial displacements of the loop itself). Flows in chromospheric structures commonly reach 3--10~km~s$^{-1}$ and can coexist with oscillatory dynamics \citep{BeckChoudhary2020_IEF, Mooroogen2017_Fibrils, Morton_2020_Superpenumbral}.

To trace the spiral flow channels, we utilise the OCCULT-2 algorithm \citep{Aschwanden_2013}, a tried-and-tested automated tracing code originally developed for coronal loop detection. Recent high-resolution observations with Solar Orbiter/EUI and SDO/AIA have revealed abundant populations of highly compact, small-scale loops with lengths of just $\sim$3--30~Mm \citep{Shrivastav2024, Madjarska2024}, validating the application of such algorithms to the smaller-scale magnetic environments observed in the chromosphere.

In this paper the extrapolation is used for topological interpretation only, as it does not capture currents, shear, or non-potential structure that may be important in and around pores \citep{WiegelmannSakurai2012_LRSP}. This context is especially useful when interpreting whether the spiral is consistent with an expanding canopy geometry or with low-lying, returning connectivity.

The paper is structured as follows. Section~\ref{sec:obs} describes the observations and data reduction, Section~\ref{sec:Methodology} details our analysis pipeline, and Section~\ref{sec:results} presents the key results. Section~\ref{sec:discussion} discusses the implications and summarises our findings.

\begin{figure*}
 \centering
 \includegraphics[width=\textwidth]{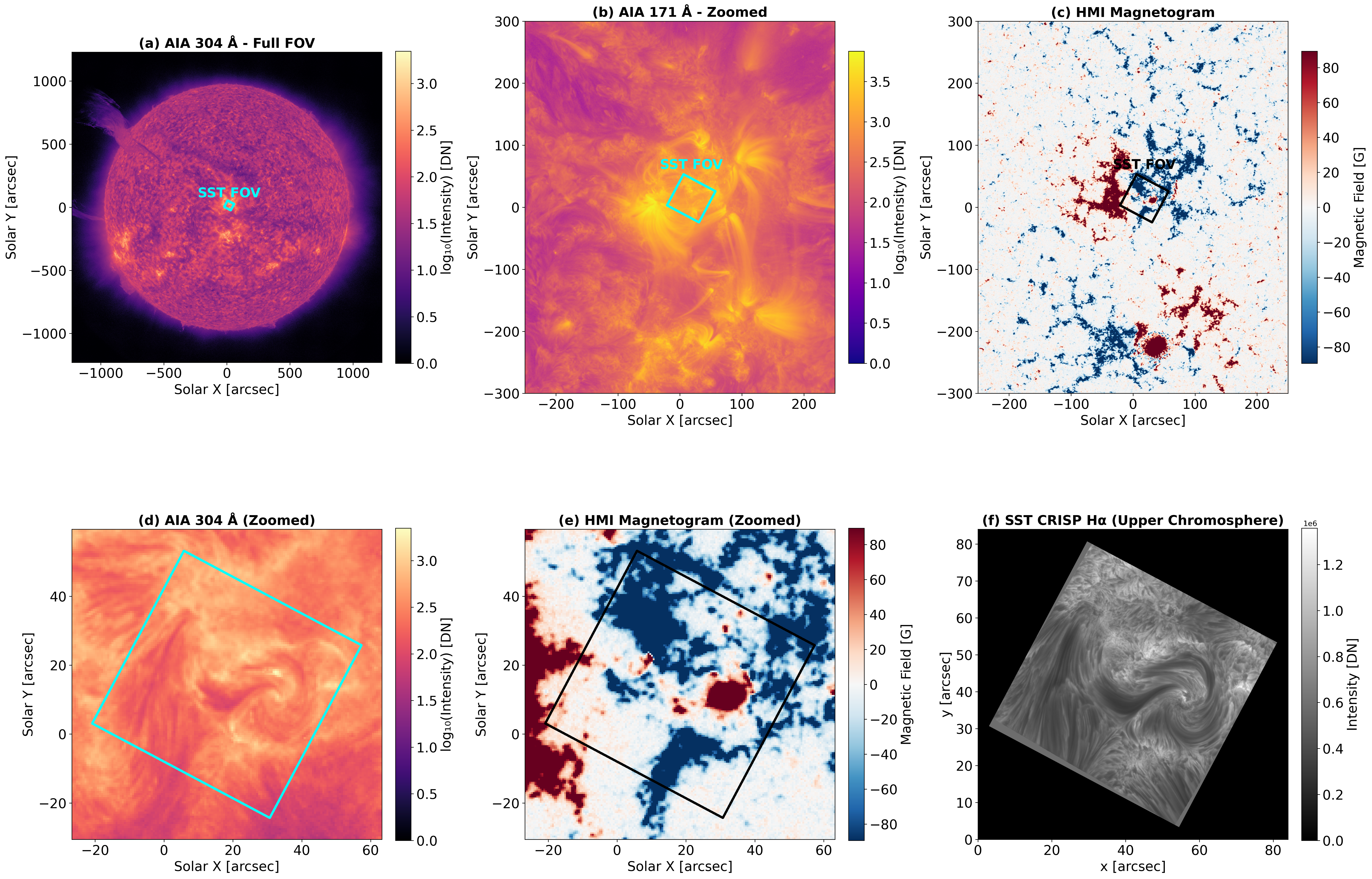}
 \caption{Multi-scale context of the giant chromospheric spiral observed on 8 October 2012. Top row: (a) AIA 304~\AA\ full-disk image showing the location of the pore near the solar equator. The cyan contour indicates the SST field-of-view (FOV). (b) AIA 171~\AA\ zoomed view revealing the large-scale trans-equatorial quadrupolar coronal loop structure. The coronal loops span both solar hemispheres. (c) HMI line-of-sight magnetogram showing the photospheric magnetic field distribution. The pore emerges within a region of opposite magnetic polarity, and thus destabilises the quadrupolar structure. Bottom row: Zoomed views with SST H$\alpha$ observation overlaid. (d) AIA 304~\AA\ chromospheric emission with the SST FOV boundary showing the chromospheric environment surrounding the pore. (e) HMI magnetogram co-aligned with the SST observations shows that the pore emerges in a region of opposite polarity. (f) SST CRISP H$\alpha$ observation of the upper chromosphere, displaying the fine-scale fibril structure and the giant spiral.}
 \label{fig:sst_observation}
\end{figure*}

\section{Observations and Data Reduction} \label{sec:obs}

The high-resolution observations analysed here were obtained with the Swedish 1-m Solar Telescope \citep[SST;][]{Scharmer_2003} using the CRisp Imaging SpectroPolarimeter \citep[CRISP;][]{Scharmer_2008}. The target was a stable magnetic pore exhibiting a prominent chromospheric spiral. The dataset was acquired on 8 October 2012 and consists of a 37-minute time series with a cadence of 9~s and an image scale of $\sim$43~km~pixel$^{-1}$. The primary diagnostic for this study is the H$\alpha$ line at 656.28~nm, sampled by CRISP at four spectral positions to capture upper-chromospheric fibrils and their dynamics. We also use a co-temporal photospheric Fe~I 630.2~nm sequence for context.

The CRISP data were reduced using the standard reduction pipeline \citep{delacruz_2015}, including dark-current subtraction, flat-fielding, and image restoration using Multi-Object Multi-Frame Blind Deconvolution. The CRISP time series was stabilised and co-aligned across the observing sequence. The CRISP coordinate system was established by aligning pores and bright points with co-temporal SDO/AIA 1700~\AA\ images, providing a stable reference and sub-SDO-pixel registration across the CRISP field of view (Figure~\ref{fig:sst_observation}). Initial data exploration and visualisation were performed using the CRISP Data Explorer \citep[CRISPEX;][]{Vissers_2012}.

For large-scale context (Figure~\ref{fig:sst_observation}), we reference full-disk observations from the Solar Dynamics Observatory (SDO). Line-of-sight magnetograms from the Helioseismic and Magnetic Imager \citep[SDO/HMI;][]{Scherrer2012} provide the photospheric magnetic environment surrounding the target region, and EUV imagery from the Atmospheric Imaging Assembly \citep[SDO/AIA;][]{Lemen2012} provides coronal context. These SDO data are used for contextual interpretation only; the quantitative tracing and oscillatory flow measurements presented in this paper are derived from the SST/CRISP time series.

%\newpage
\section{Methodology}\label{sec:Methodology}

\begin{figure}[H]
\centering
\resizebox{\columnwidth}{!}{%
\begin{tikzpicture}[
 node distance=5mm,
 every node/.style={font=\small}
]
\tikzset{
 stage/.style={
 rectangle, rounded corners=2mm, draw=black, fill=gray!6,
 line width=0.7pt, text width=0.92\columnwidth, align=left, inner sep=5pt
 },
 connector/.style={-Stealth, line width=0.8pt},
 output/.style={font=\small\itshape, text=blue!60!black}
}

% ---- S1
\node[stage] (s1) {\textbf{S1. Input data processing}\\
\emph{Input:} SST FITS files\\
\emph{Process:} Convert from data cubes, apply solar derotation to stabilize the central pore.\\
{\color{blue!60!black}$\rightarrow$ Output: Derotated FITS cubes}};

% ---- S2
\node[stage, below=of s1] (s2) {\textbf{S2. Automated Loop Detection (OCCULT2)}\\
Trace the edges of plasma flows by following intensity gradients. Run on both original and inverted images to detect bright and dark flows.\\
{\color{blue!60!black}$\rightarrow$ Output: Flow channel coordinates \& XT-cuts}};
\draw[connector] (s1) -- (s2);

% ---- S3
\node[stage, below=of s2] (s3) {\textbf{S3. Multi Stage Loop Filtering}\\
Remove duplicates from parameter sweeps; filter by size ($>$8\,Mm); apply overlap-based filters (0\% and $>$70\% overlap).\\
{\color{blue!60!black}$\rightarrow$ Output: Filtered flow channel sets}};
\draw[connector] (s2) -- (s3);

% ---- S4
\node[stage, below=of s3] (s4) {\textbf{S4. Statistical Properties of Detected Loops}\\
Statistical analysis is performed on the detected loops.\\
{\color{blue!60!black}$\rightarrow$ Output: Statistical analysis outputs}};
\draw[connector] (s3) -- (s4);

% ---- S5
\node[stage, below=of s4] (s5) {\textbf{S5. Loop Evolution Tracking}\\
Associate loops across consecutive frames using a custom algorithm that group matches and assign unique evolution IDs.\\
{\color{blue!60!black}$\rightarrow$ Output: Tracked flows (evolution IDs)}};

% Branch line from S3 to S5 (bypasses S4 without crossing it)
\draw[connector]
  (s3.south) -- ++(0,-3mm)
  -| ([xshift=-0.7cm]s4.west |- s5.west)
  -- (s5.west);

% ---- S6
\node[stage, below=of s5] (s6) {\textbf{S6. Automated Thread detection}\\
Detect fine-scale oscillating threads in space-time (XT) cuts using a custom Python algorithm.\\
{\color{blue!60!black}$\rightarrow$ Output: Thread coordinates (\texttt{.npz})}};
\draw[connector] (s5) -- (s6);

% ---- S7
\node[stage, below=of s6] (s7) {\textbf{S7. Oscillatory mode fitting \& analysis}\\
Fit models to thread trajectories to extract amplitude, period/frequency, and phase of oscillations.\\
{\color{blue!60!black}$\rightarrow$ Output: Oscillation parameters}(\texttt{.npz}/CSV)};
\draw[connector] (s6) -- (s7);

% ---- S8
\node[stage, below=of s7] (s8) {\textbf{S8. Visualisation}\\
Create maps correlating oscillation properties with flow channel geometry (curvature, length, evolution ID).\\
{\color{blue!60!black}$\rightarrow$ Output: Science-ready plots}};
\draw[connector] (s7) -- (s8);

\end{tikzpicture}%
}
\caption{The end-to-end data analysis pipeline.}
\label{fig:pipeline_tikz}
\end{figure}
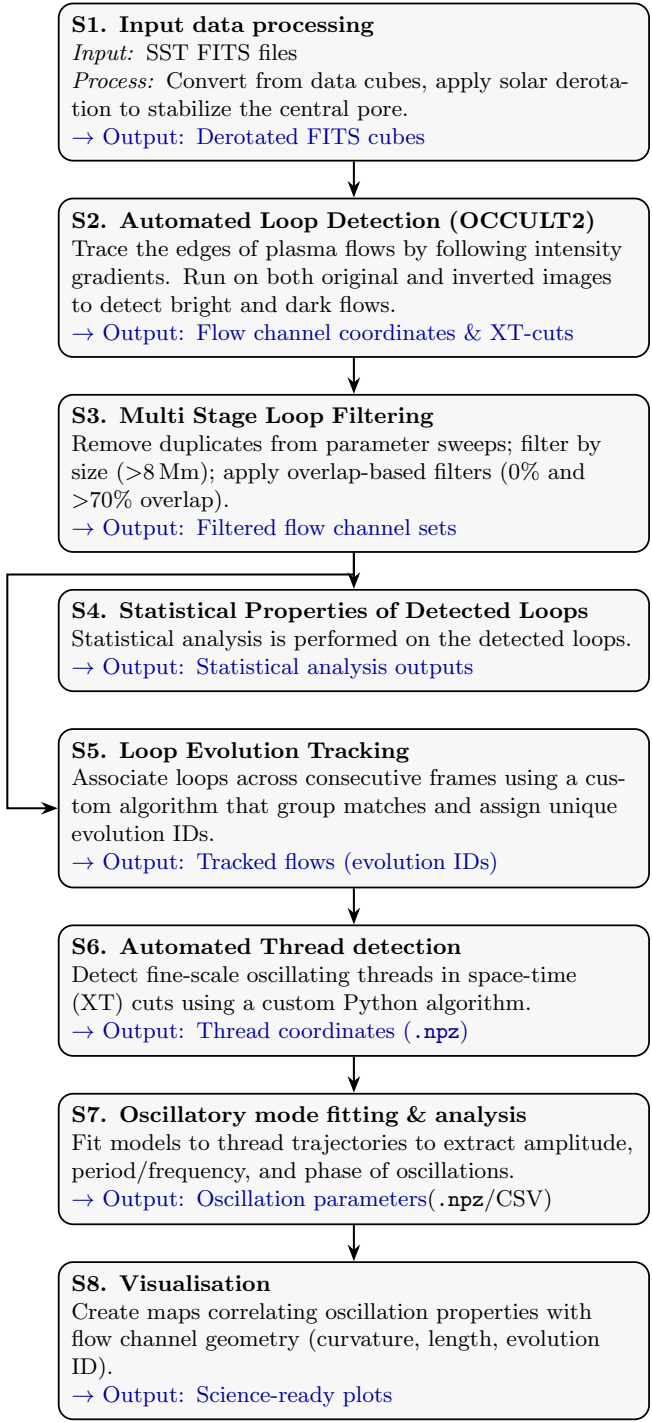

We process the high cadence SST observation through a multi-stage pipeline to identify the channels of plasma flows that form the spiral in the solar chromosphere. We also track the evolution of these loops and quantify the properties of plasma flows inside these loops. The pipeline is depicted in Figure \ref{fig:pipeline_tikz}. Each stage (S1--S8) transforms the data progressively from raw cubes to FITS to final scientific insights. This section provides an overview of each stage. Complete algorithmic details for the thread detection and oscillatory mode fitting procedures are provided in a companion methodology paper \citep{SaneshwarInPrep}.

To ensure clarity throughout this paper, we explicitly define the terminology used to describe the features analysed in this study. ``Loops'' refer to the magnetic flux tubes that anchor the spiral, along which plasma moves outward from the pore. A single, brief injection of plasma is classed as a ``pulse''. When plasma flows are continuous and sustained enough to form a distinct, trackable structure in the space--time (XT) diagram extracted along a loop, we refer to this feature as a ``thread''. A thread may therefore represent a flow, an oscillation, or a combination of both, as they appear projected in the XT cut. Throughout this paper, references to ``flows'' denote oscillatory plasma motions along the loop threads, which may manifest as isolated pulses or as repetitive, periodic displacements; this oscillatory behaviour is what motivates the multi-component mode analysis.

\subsection{Input data processing (S1)}\label{subsec:input_derotation}
The observational data, originally stored in the telescope's proprietary FITS cube format (a multi-dimensional array storing spectral, spatial, and temporal data in a single file), are first converted to a series of FITS files (one file per timestep). A crucial preprocessing step is the correction for solar differential rotation. The CRISP time series showed apparent motion of the central pore, which would corrupt tracking analysis. We developed and applied a derotation algorithm that tracks the pore location and applies differential rotation corrections to keep the pore fixed in the frame of reference throughout the observation so that when we extract the XT cuts of the loops there will be zero differential flows arising which would lead to misinterpretation of the plasma dynamics in the spiral arms.

\subsection{Automated Loop Detection (OCCULT2) (S2)}\label{subsec:occult}
We use the Oriented Coronal Curved Loop Tracing (OCCULT2) algorithm \citep{Aschwanden_2013} to automatically trace the spiral arm flow channels by following intensity gradients. Since the spiral contains both bright and dark flow channels, we apply detection algorithm on both the original H$\alpha$ images and their inverted counterparts. We perform a parameter sweep over smoothing scales, curvature thresholds, and intensity cutoffs to ensure detection across varying seeing conditions.

\subsection{Multi-Stage Loop Filtering (S3)}\label{subsec:filtering}
The parameter sweep produces many redundant detections. We apply a four-stage filtering pipeline: (1) duplicate removal from parameter sweeps, (2) removal of duplicate loops after concatenation of detections from original and inverted images, (3) size filtering to remove features shorter than 8 Mm in order to avoid jets being studied as loops, and (4) overlap filtering to create two catalogues: a unique set with no overlapping loops, and a partial-overlap set retaining loops with $>70\%$ overlap. The partial overlap filtering was done in order to get a catalogue of loops that are oscillating. A threshold of $>70\%$ was chosen as it provides an optimal balance between temporal continuity and morphological flexibility: a lower threshold (e.g. $>65\%$) risks including loop pairs that are not genuinely the same evolving structure but merely spatially coincident detections, while a higher threshold would be overly restrictive and fail to capture loops that undergo real morphological evolution between frames. At $>70\%$, the retained loops show sufficient spatial consistency to be confidently identified as the same structure across consecutive timesteps, while still accommodating the natural evolution of the loop geometry over time. The outcome of this filtering process is illustrated in Figure \ref{fig:filtering}.

\subsection{Statistical Properties of Detected Loops (S4)}\label{subsec:statistics}
We analysed the statistical properties of the final filtered loop catalogue. To understand the geometry of each loop, we calculated its curvature, $\kappa$. For a two-dimensional curve parametrised as $\mathbf{r}(s) = (x(s), y(s))$, where $s$ is pixel index number, the curvature is defined as
\begin{equation}
\kappa(s) = \frac{|x'(s) y''(s) - y'(s) x''(s)|}{\big(x'(s)^2 + y'(s)^2\big)^{3/2}} ,
\end{equation}
with primes denoting derivatives with respect to $s$. This provides us with a per-pixel curvature profile for each loop, from which we computed the mean curvature as a single representative value. The relationship between loop length and mean curvature is shown in Figure \ref{fig:length_vs_curvature}.

\subsection{Loop Evolution Tracking (S5)}\label{subsec:evolution}
To track the temporal evolution of the spiral structure, we associate loops across consecutive frames using a custom matching algorithm that assigns unique evolution IDs to loops with lifetimes over multiple frames. A loop is given a unique evolution ID and the algorithm checks if another loop exists at similar coordinates in the consecutive time frame and assigns that loop that same evolution ID so we have a catalogue of loops that evolve in time. 

\subsection{Automated Thread Detection (S6)}\label{subsec:thread_detection_xt}
An XT cut is taken of each of these loops. In these XT cuts we see how plasma behaves along the loop over time. We observe oscillatory patterns in the field-aligned loop flows in these XT cuts and to study the properties of these features, we developed a custom algorithm to quantify the properties of these features. We call these oscillatory features threads and we identify fine-scale oscillating threads within the space-time (XT) cuts extracted along each detected loop using a three-stage pipeline. First, we enhance contrast and suppress noise using Contrast Limited Adaptive Histogram Equalization \citep{Zuiderveld1994_CLAHE} and median filtering. Second, we extract thread boundaries using Canny edge detection \citep{Canny1986}. Third, we link these edge pixels across time using graph-based dynamic programming that minimizes a cost function favouring smooth, continuous paths while allowing small temporal gaps to handle bad seeing. We restrict the minimum thread length to be at least 10 frames. This is to avoid tracking small jets. An example of the thread detection output is shown in Figure \ref{fig:xt_cut_side_by_side}.

\subsection{Oscillatory Mode Fitting and Analysis (S7)}\label{subsec:fitting}
We measure oscillations in the detected threads using an automated multi-component oscillatory mode fitting algorithm. Before fitting, the code checks whether a thread is oscillatory or if it is just a short-lived transient (a pulse). If the thread contains fewer than $\sim$1.5--2 cycles over its lifetime, it is treated as a pulse/dip and we fit a Gaussian pulse/dip model to estimate an amplitude and a duration (using half-maximum edge times), rather than forcing a multi-component oscillation model. All remaining threads are treated as oscillatory. After detrending to remove slow drifts, we fit each oscillatory thread with up to three exponentially damped cosines:
\begin{equation}
x_{\mathrm{fit}}(t) = \sum_{i=1}^{N} A_i \, e^{-\gamma_i t}\cos\!\big(2\pi f_i t + \phi_i\big),
\label{eq:multiwave}
\end{equation}
where $A_i$ is amplitude, $\gamma_i$ is damping rate, $f_i$ is frequency (period $P_i=1/f_i$), and $\phi_i$ is phase. We use FFT with a Hann window \citep{Harris1978} to seed initial frequency guesses within the period band [1.5, 15] minutes. The code then performs multiple random restarts to find the best-fit parameters. Each input parameter combination is tested to find a fit that avoids overfitting and best describes the oscillation. Model selection uses both the Akaike Information Criterion \citep[AIC;][]{Akaike1974} and the Bayesian Information Criterion \citep[BIC;][]{Schwarz1978}; we only accept additional components if both criteria improve by $\geq 4$ compared to simpler models. Residual diagnostics (Kolmogorov-Smirnov, Anderson-Darling, D'Agostino $K^2$, Ljung-Box tests) assess fit quality. Examples of 1-, 2-, and 3-component fits are shown in Figures \ref{fig:thread29}--\ref{fig:thread62_comparison}.

\section{Results} \label{sec:results}

\begin{figure*}
\centering
\includegraphics[width=2.0\columnwidth]{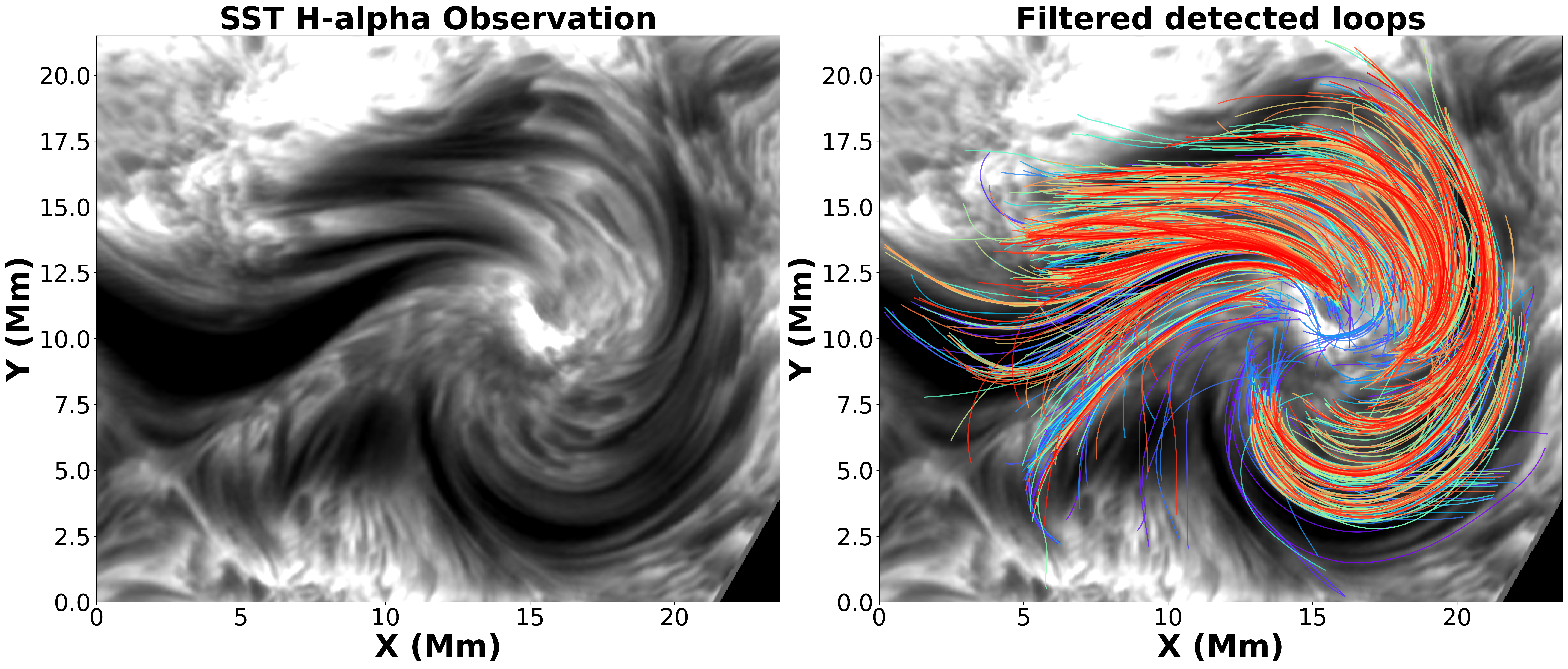}
\caption{Illustration of the detected loops after filtering. Left: The SST observation. Right: The unique loop catalogue produced after the four-stage filtering pipeline (step S3, Figure~\ref{fig:pipeline_tikz}), retaining loops with no spatial overlap to avoid redundant detections. Colours are random and do not represent any information. The grayscale in the panels represents $H\alpha$ intensity in detector counts (DN).}
\label{fig:filtering}
\end{figure*}

\begin{figure*}
 \centering
 \includegraphics[width=1.5\columnwidth]{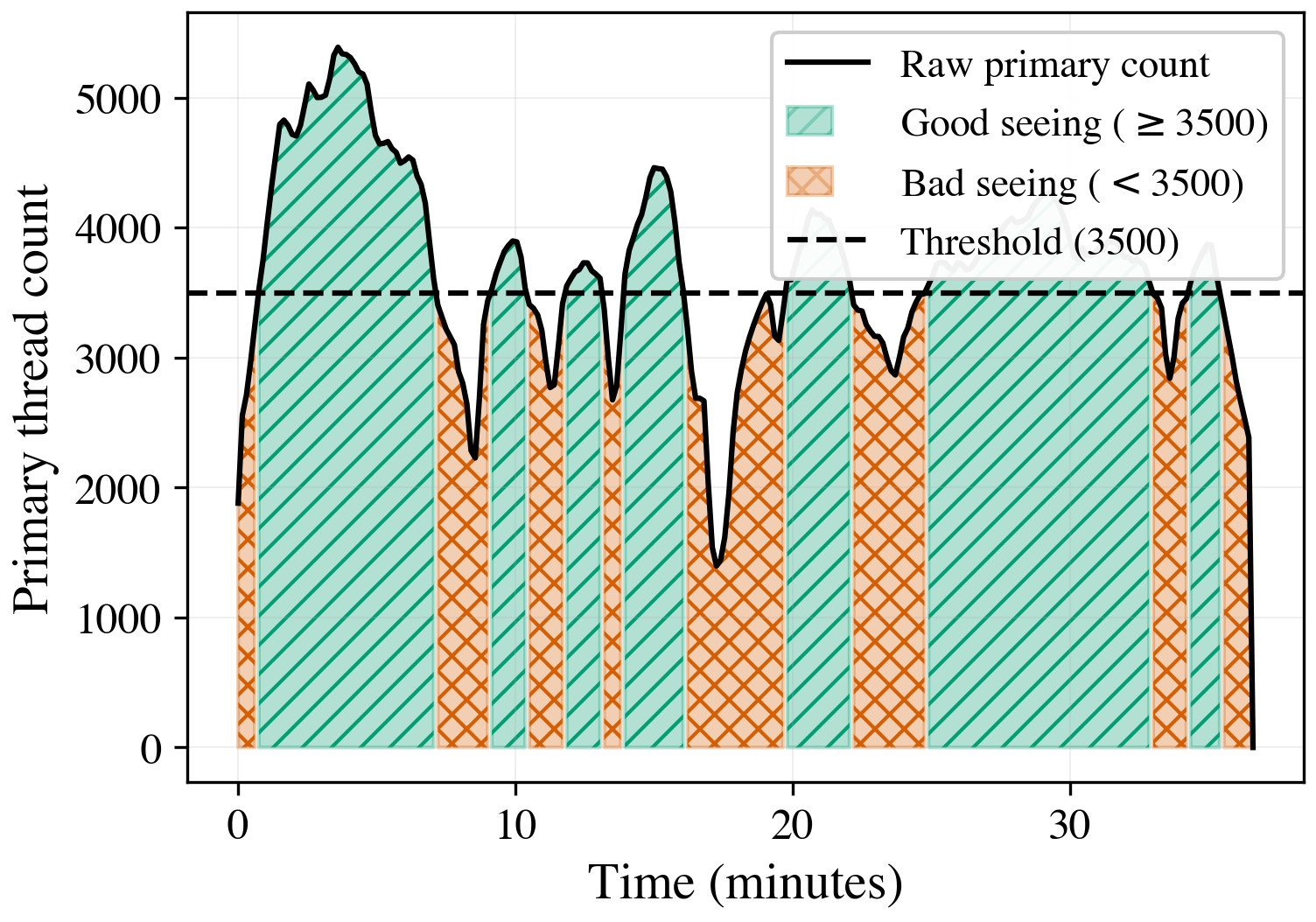}
 \caption{Atmospheric seeing quality diagnostic based on primary thread detection counts. Green shading indicates timesteps with good seeing ($\geq 3500$ threads) that were retained for analysis, while red shading shows timesteps with poor seeing ($< 3500$ threads) that were excluded.}
 \label{fig:seeing_quality}
\end{figure*}

The automated loop detection and filtering pipeline (S2-S3) successfully identified and catalogued loops that constitute the spiral structure. Figure \ref{fig:filtering} shows the output of detected loops after our multi-stage filtering process. The observation drops seeing quality at times during the entire observation and it is correlated to drop in thread detection. To remove any effects of bad seeing on our data we filter out the time periods where the count of detected threads drops below 3500. This specific threshold was selected to isolate the peaks in detection counts—which directly correlate with periods of optimal seeing. Figure \ref{fig:seeing_quality} shows the drop in the thread detection count. In total we drop 66 timesteps and thus focus on the threads in the remaining time periods.

\subsection{Statistical properties of detected loops}
\begin{figure}[!htbp]
 \centering
 \includegraphics[width=1.0\columnwidth]{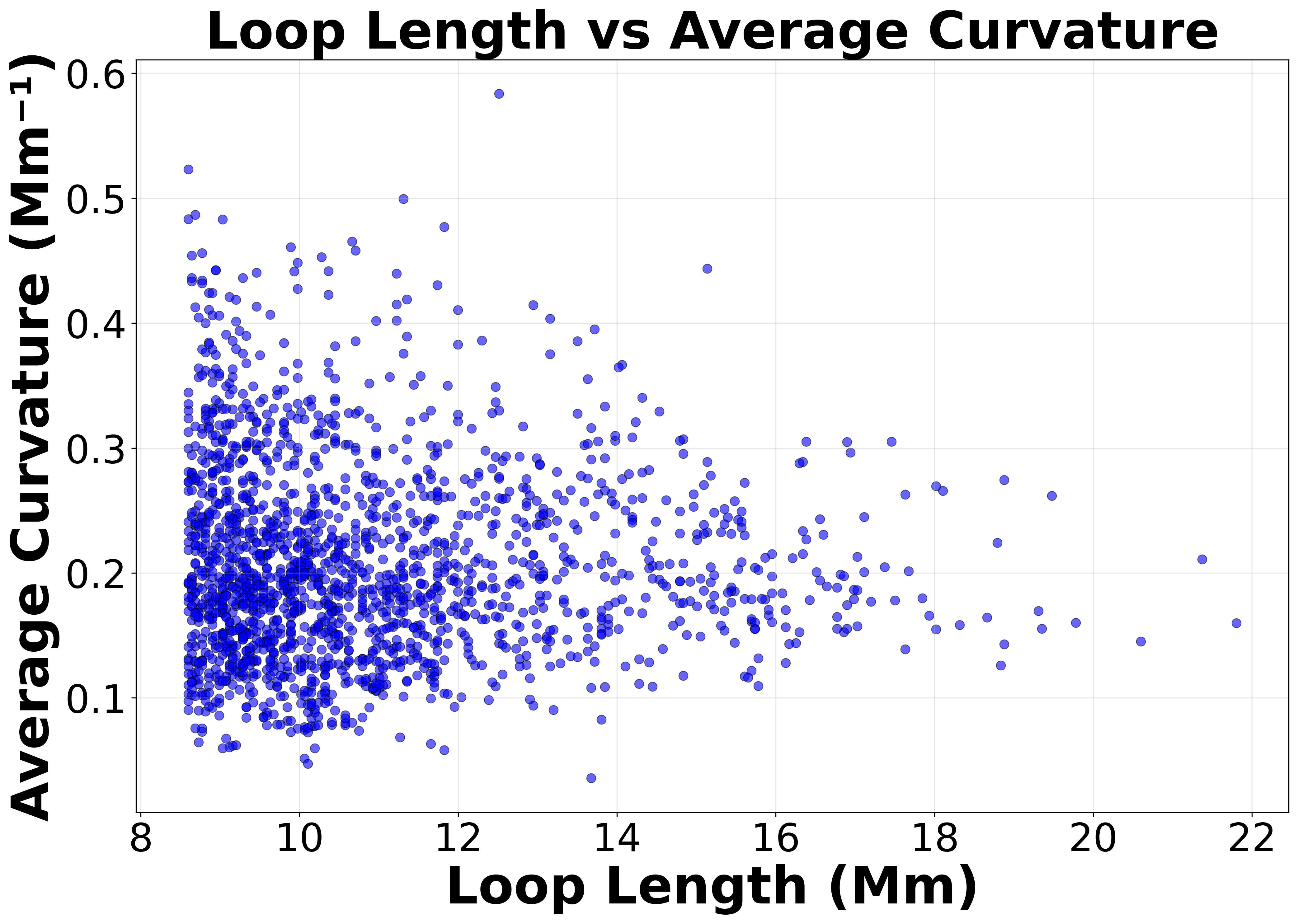}
 \caption{Scatter plot of mean loop curvature versus loop length. Obtained after step 4 of the analysis pipeline.}
 \label{fig:length_vs_curvature}
\end{figure}

The geometric analysis of the filtered loop catalogue reveals an important property of our detection algorithm. Figure \ref{fig:length_vs_curvature} shows a scatter plot of the mean curvature of each loop versus its length. The Pearson correlation coefficient between these two properties is $-0.029$. This lack of correlation demonstrates that the algorithm does not impose a simple morphological bias: the length of a detected loop is not a predictor of its shape. Long loops are just as likely to be straight as they are to be highly curved, indicating that the detection is sensitive to a wide variety of loop morphologies.

\subsection{Thread Detection in Space-Time Cuts}
\begin{figure*}[t!]
\centering
\begin{subfigure}{0.48\textwidth}
 \centering
 \includegraphics[width=\linewidth]{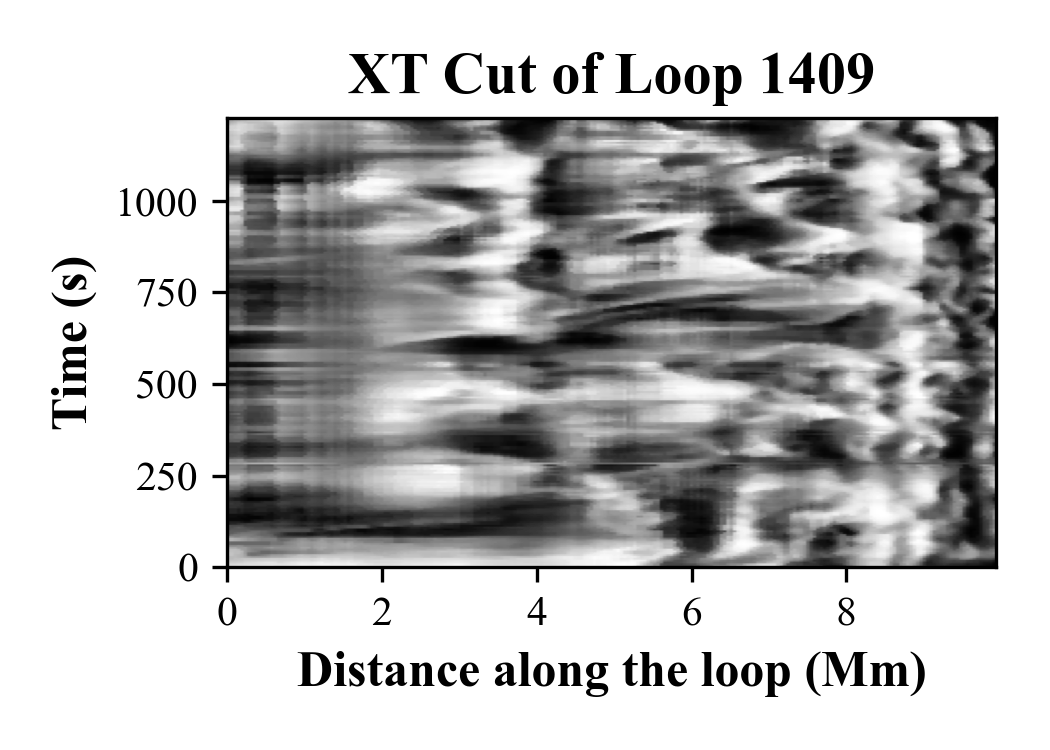}
 \caption{Loop 1409 XT cut, after preprocessing.}
 \label{fig:xt_processed}
\end{subfigure}
\hfill
\begin{subfigure}{0.48\textwidth}
 \centering
 \includegraphics[width=\linewidth]{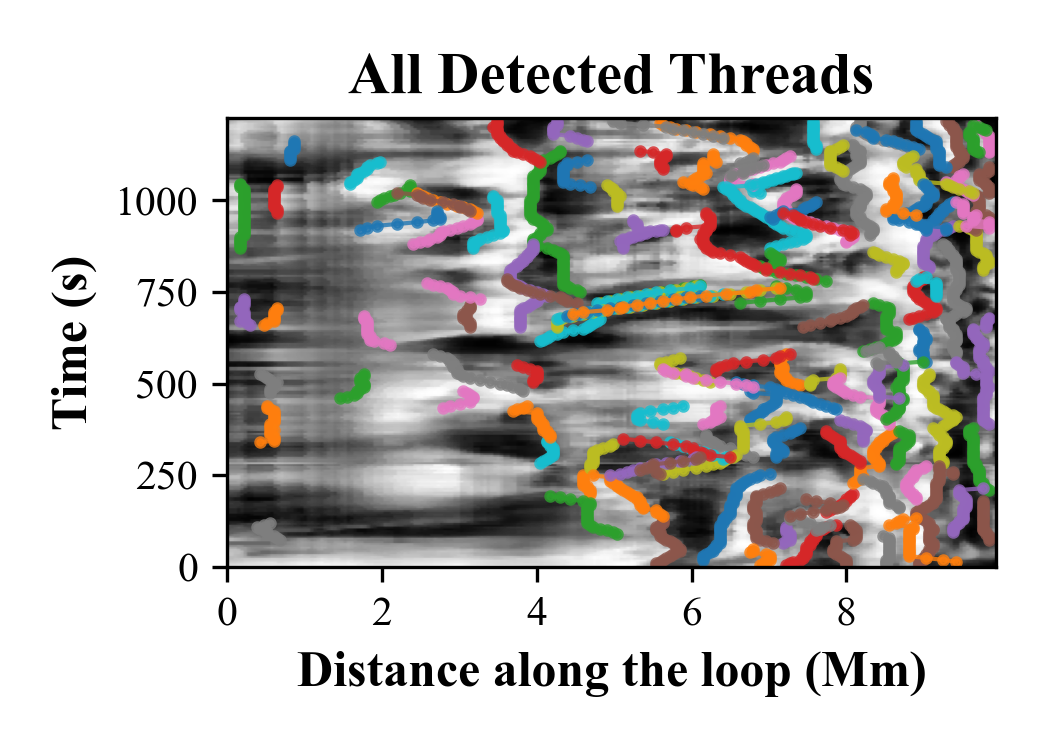} 
 \caption{Detected threads overlaid on the same XT cut. }
 \label{fig:xt_threads_detected}
\end{subfigure}
\caption{Example of the thread detection pipeline output after S6 for Loop 1409. Panel (a) shows the contrast-enhanced, denoised XT diagram. Panel (b) shows the output after edge extraction and graph tracking; 132 threads were found. The grayscale in the panels represents $H\alpha$ intensity in detector counts (DN).}
\label{fig:xt_cut_side_by_side}
\end{figure*}

The automated thread detection algorithm successfully identified fine-scale oscillating features within the detected loops. Figure \ref{fig:xt_cut_side_by_side} demonstrates the thread detection pipeline applied to Loop 1409. Panel (a) shows the contrast-enhanced, denoised XT diagram after preprocessing, where the flow structures are clearly visible in space-time. Panel (b) shows the output after edge extraction and graph tracking, where 132 individual threads were successfully identified and tracked through the observation. It is worth noting that threads are more prevalent near the loop footpoints, as this is where oscillatory patterns are most prominent.

\subsection{Oscillatory Mode Fitting Examples}

\begin{figure*}[!htbp]
 \centering
 % Top row: 1-component and 2-component fits
 \begin{subfigure}{0.5\textwidth}
  \centering
  \includegraphics[width=\linewidth]{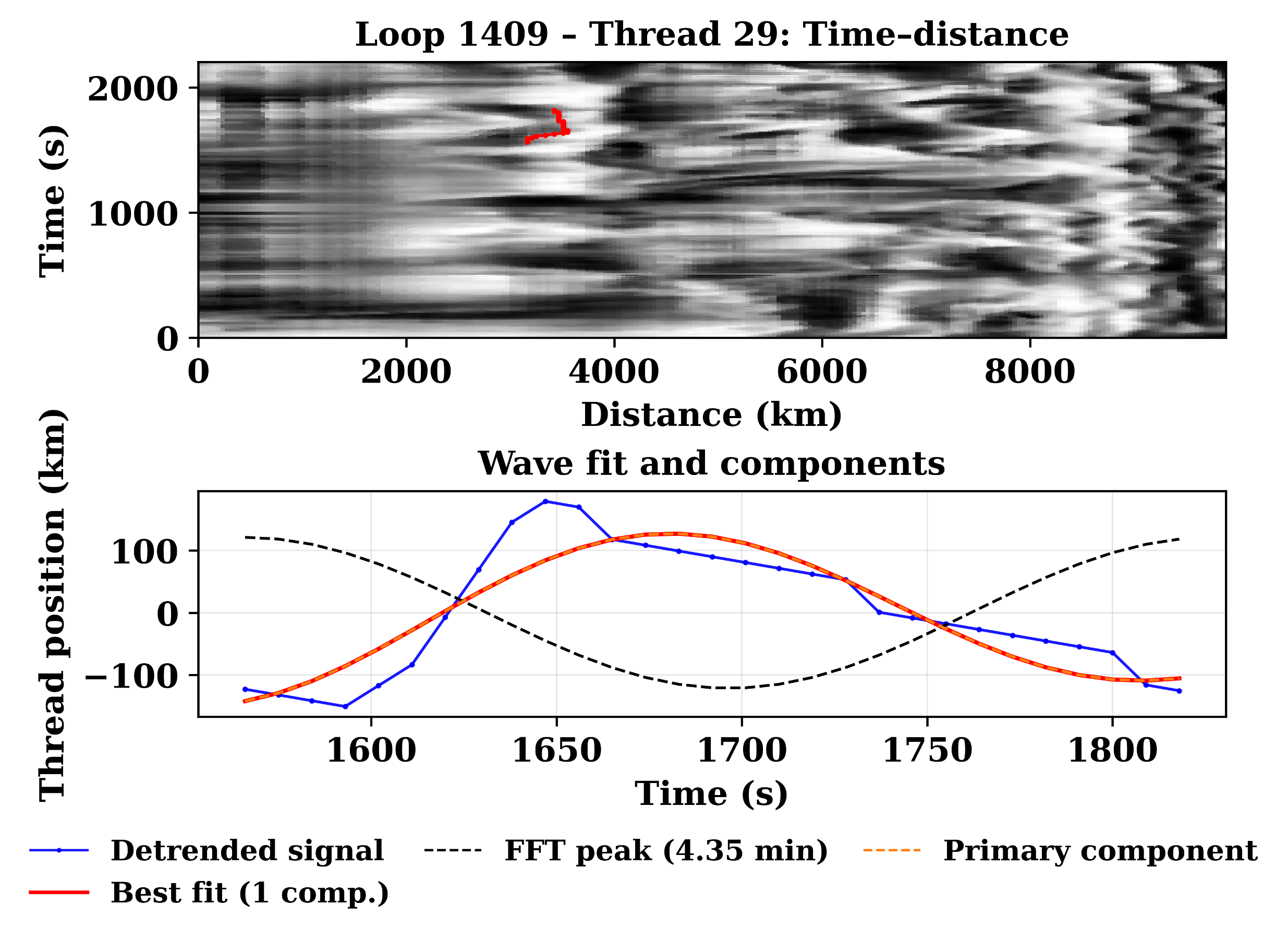}
  \caption{1-component fit (thread 29)}
  \label{fig:thread29}
 \end{subfigure}
 \hfill
 \begin{subfigure}{0.49\textwidth}
  \centering
  \includegraphics[width=\linewidth]{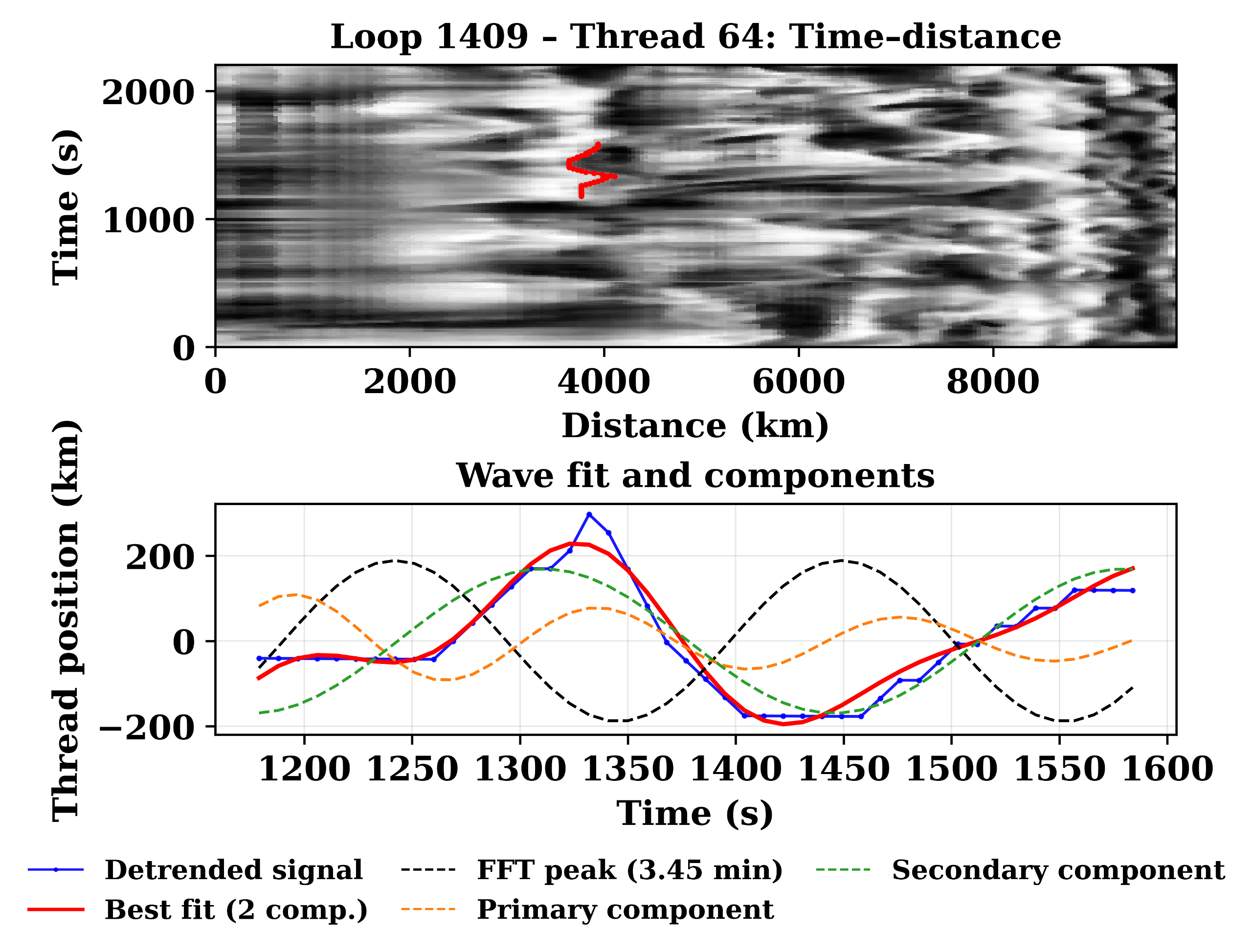}
  \caption{2-component fit (thread 48)}
  \label{fig:thread64}
 \end{subfigure}
 
 \vspace{0.5em}
 
 % Bottom row: Both panels for 3-component fit
 \begin{subfigure}{0.50\textwidth}
  \centering
  \includegraphics[width=\linewidth]{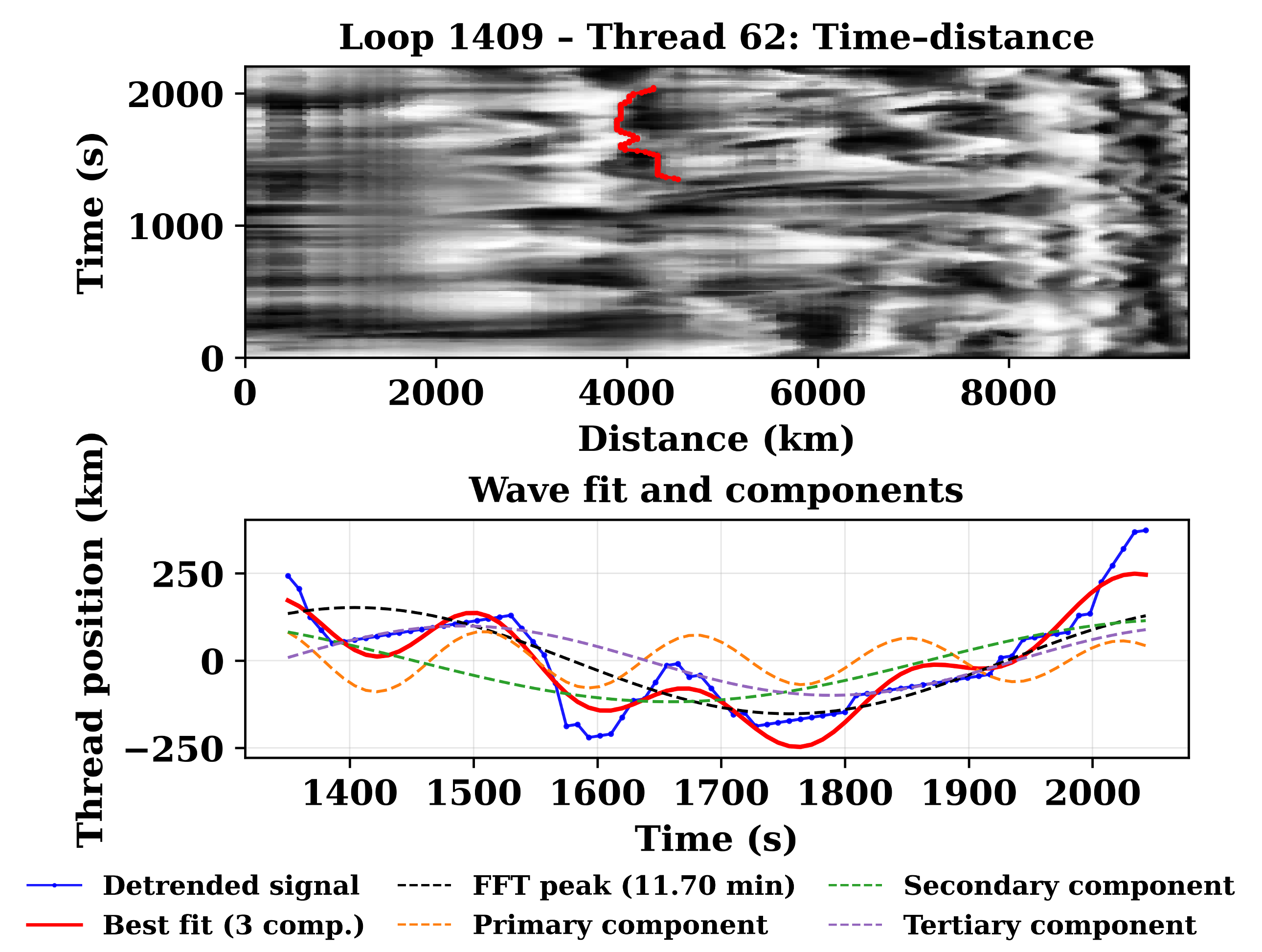}
  \caption{3-component fit (thread 62)}
  \label{fig:thread62_analysis}
 \end{subfigure}
 \hfill
 \begin{subfigure}{0.48\textwidth}
  \centering
  \includegraphics[width=\linewidth]{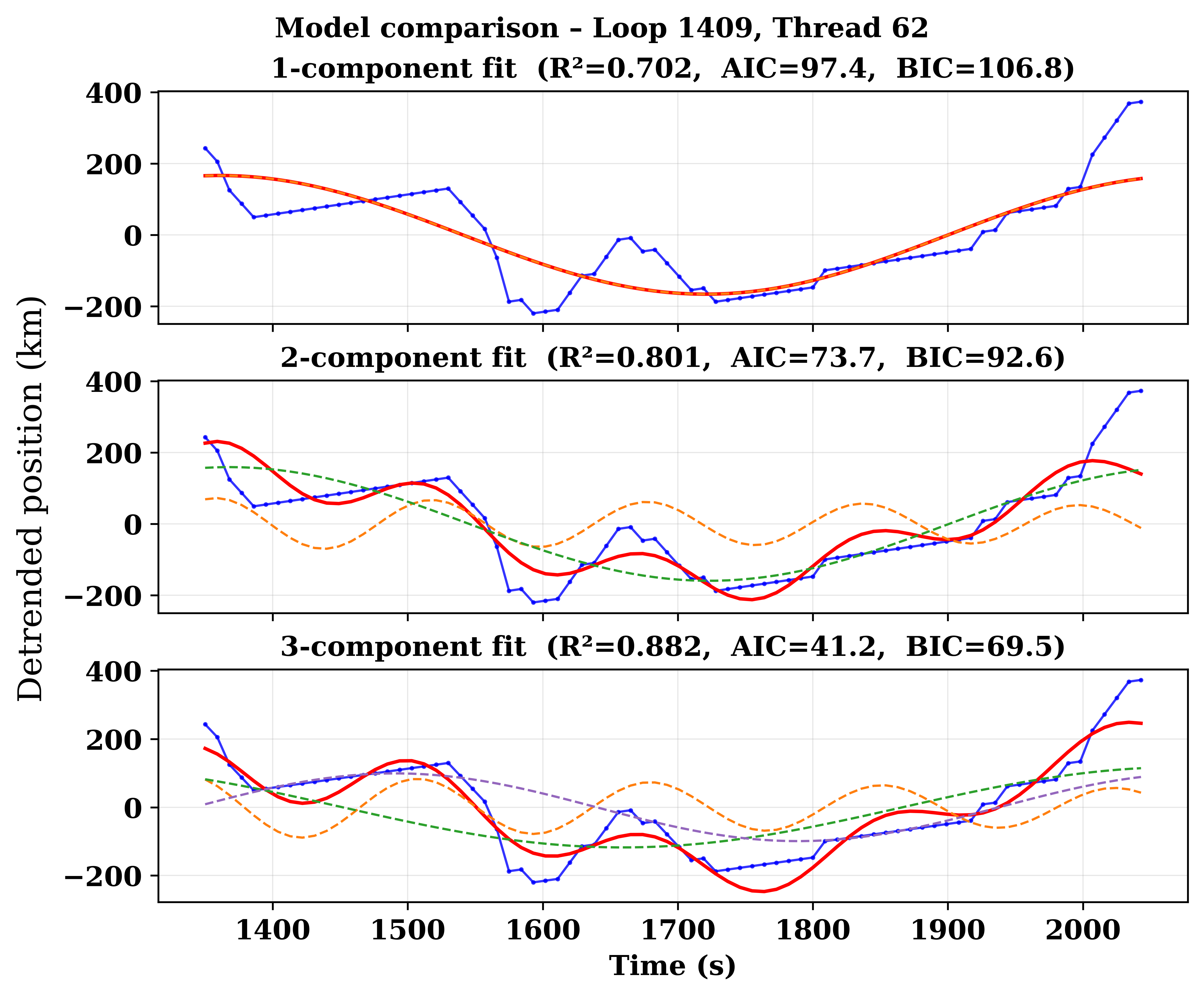}
  \caption{Model comparison for thread 62}
  \label{fig:thread62_comparison}
 \end{subfigure}
 
 \caption{Multi-component fitting examples for loop in Figure \ref{fig:xt_cut_side_by_side} after step S7 of the analysis pipeline. (a) 1-component Pulse fit showing detrended data and model for thread 29. (b) 2-component fit for thread 64 demonstrating the need for an additional component. (c) 3-component fit for thread 62 with detrended data and model. (d) Model comparison panel for thread 62. The grayscale in the panels represents $H\alpha$ intensity in detector counts (DN).}
 \label{fig:multicomponent_fits}
\end{figure*}

The multi-component fitting algorithm was applied to all detected threads to extract oscillation properties. Figure~\ref{fig:multicomponent_fits} shows representative examples: thread~29 (a pulse-like event), thread~64 (a clear 2-component case), and thread~62 (a clear 3-component case). The recovered parameters for these examples are summarised in Table~\ref{tab:fitting_cases_extended}.

Thread~29 (Figure~\ref{fig:thread29}) is an example of the code identifying a pulse-like signal and then finding the best single-component solution through multiple random restarts (50 iterations). Out of those iterations, the final selected fit is the one with the lowest AIC and BIC, i.e. the best trade-off between matching the signal and not overfitting. For this thread, the algorithm recovers a primary amplitude of 991.67~km and a period of 4.22~min, with a damping time of 13.6~min (Table~\ref{tab:fitting_cases_extended}; AIC = 4.2, BIC = 9.7).

Thread~64 (Figure~\ref{fig:thread64}) shows a case where one component is not enough. The algorithm selects a 2-component fit because adding a second component significantly improves the information criteria, whereas adding a third component is not justified (i.e. it would start fitting noise rather than real structure). In the selected 2-component model, the primary component has an amplitude of 1888.85~km and a period of 2.35~min (damping time 6.99~min). The secondary component captures a weaker longer-period oscillation (169.34~km, 4.48~min), with an effectively undamped behaviour over the fitted duration ($\tau \rightarrow \infty$) (Table~\ref{tab:fitting_cases_extended}).

Thread~62 (Figure~\ref{fig:thread62_analysis}) is the strongest example of a 3-component fit. The model comparison panel (Figure~\ref{fig:thread62_comparison}) shows that as components are added the fit quality improves and the AIC/BIC values drop, so the extra components are actually warranted. The recovered components include a long-period primary signal (13.82~min; 117.74~km; $\tau \rightarrow \infty$), plus two shorter-period components at 5.11~min (134.11~km; damping time 4.25~min) and 3.20~min (83.40~km; damping time 8.13~min) (Table~\ref{tab:fitting_cases_extended}). This is basically the cleanest demonstration of why the multi-component approach is needed in some threads, and how the AIC/BIC criteria stop the model from adding extra components unless they are statistically justified.

Table \ref{tab:fitting_cases_extended} summarizes the parameters recovered from these example fits.

\begin{deluxetable}{lrrrrrrr}
\tablecaption{\textbf{Summary of the multi-component oscillatory mode fitting for the case study threads}.
\label{tab:fitting_cases_extended}}
\tablehead{
\colhead{Case Study} & \colhead{Component} & \colhead{Amplitude} & \colhead{Period} & \colhead{Damping Time} & \colhead{$R^2$} & \colhead{AIC} & \colhead{BIC} \\
\colhead{} & \colhead{} & \colhead{(km)} & \colhead{(min)} & \colhead{(min)} & \colhead{} & \colhead{} & \colhead{}
}
\startdata
Thread 29 & Primary & 991.67 & 4.22 & 13.6 & 0.837 & 4.2 & 9.7 \\
\hline
Thread 64 & Primary & 1888.85 & 2.35 & 6.99 & 0.764 & 38.1 & 45.4\\
Thread 64 & Secondary & 169.34 & 4.48 & $\infty$ & 0.953 & -28.5 & -13.9 \\
\hline
Thread 62 & Primary & 117.74 & 13.82 & $\infty$ & 0.701 & 97.4 & 106.8 \\
Thread 62 & Secondary & 134.11 & 5.11 & 4.25 & 0.801 & 73.73 & 92.6\\
Thread 62 & Tertiary & 83.40 & 3.20 & 8.13 & 0.881 & 41.2 & 69.5 \\
\enddata
\end{deluxetable}

\subsection{Intensity distribution of loops and threads vs background}
To assess the energetic impact of these flows, we analysed the H$\alpha$ intensity distributions of the detected features. We categorised pixels into three populations: the general background (pixels that are not part of any loops or threads), the Loop-Only pixels (loop pixels with no detected oscillation), and Thread pixels (loop pixels of continuously tracked intensity features).

\begin{figure*}[h]
  \centering
  \begin{subfigure}{0.48\textwidth}
    \centering
    \includegraphics[width=\linewidth]{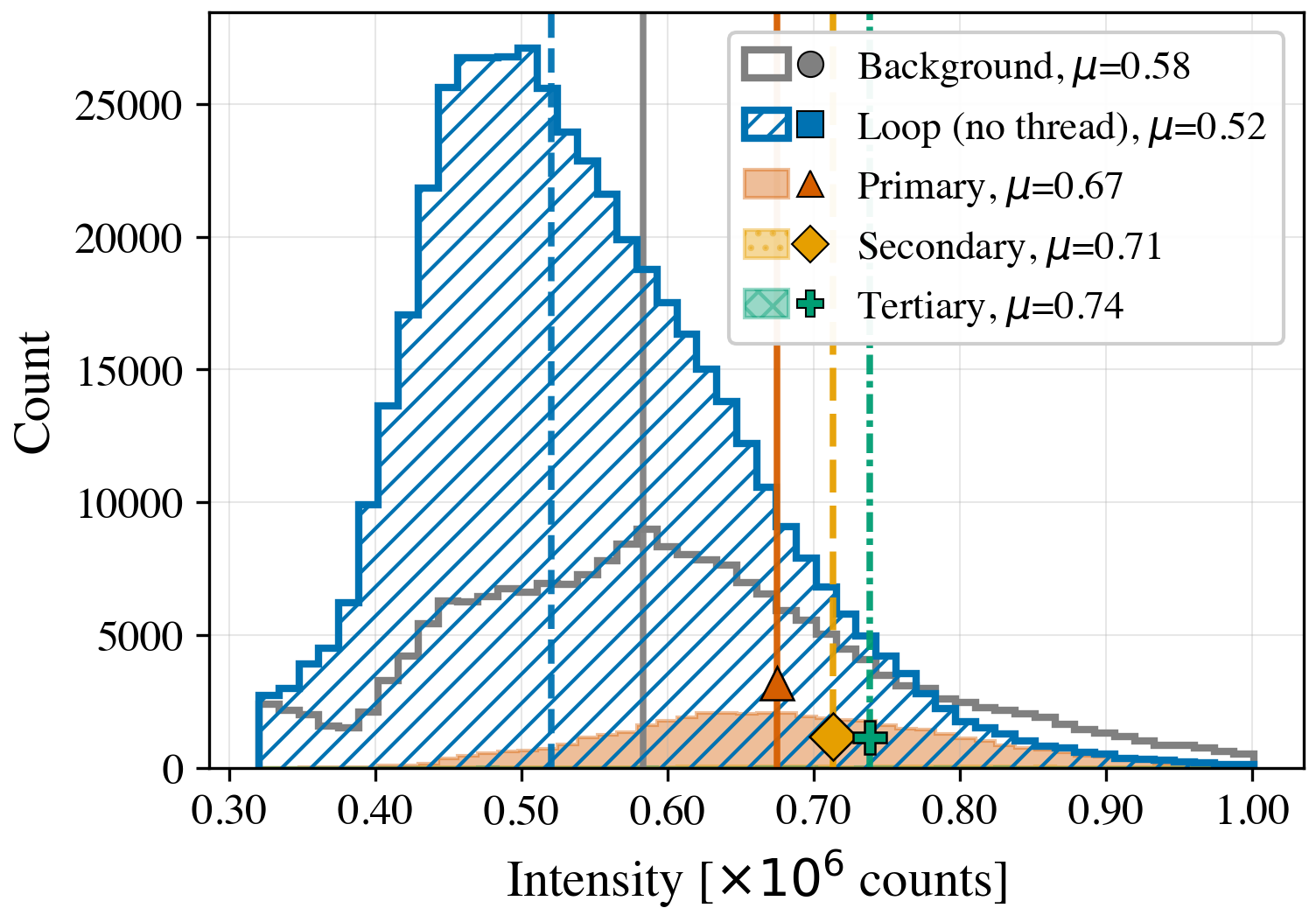}
    \caption{Scatter intensity distribution of all pixels in all regions}
    \label{fig:intensity_raw}
  \end{subfigure}
 \vspace{0.5cm}
 \begin{subfigure}{\columnwidth}
 \centering
 \includegraphics[width=\linewidth]{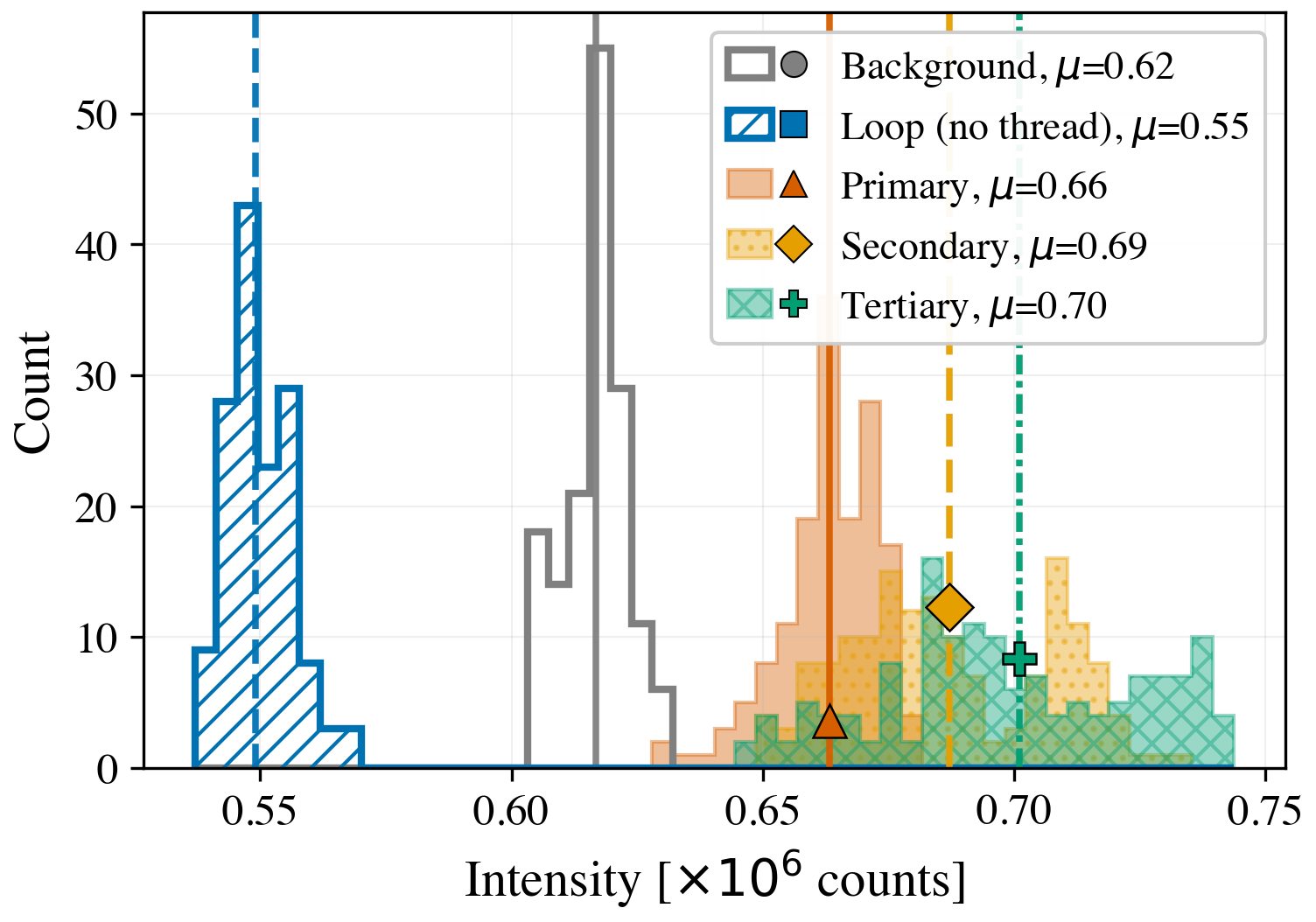}
 \caption{Distribution of time-step averaged intensity}
 \label{fig:intensity_filtered}
 \end{subfigure}
 \caption{Intensity signatures of the spiral structure. (a) Aggregate pixel distribution. The distribution of raw pixel intensities across the entire dataset, showing the broader spread of values. (b) Temporal averaged intensity. This histogram displays the distribution of the averaged intensity calculated for each category at every timestep. By averaging all pixels within a category for each frame.}
\label{fig:thermal_analysis}
\end{figure*}

 Figure~\ref{fig:thermal_analysis} presents the intensity distributions for these categories. Panel (a) shows the raw pixel intensity distribution across the full dataset, while panel (b) shows the distribution of the mean intensity per timestep (i.e. for each frame we average all pixels within a category and then plot the distribution of those timestep averages). Together, these plots reveal a clear intensity separation between loops, threads, and the background:

\begin{enumerate}
 \item \textbf{Loops (cool/absorbing):} The Loop Only population sits at the lowest intensities in both panels. In the raw distribution (Figure~\ref{fig:thermal_analysis}a) it peaks at $\sim 5.8 \times 10^5$ counts, and in the timestep-averaged distribution (Figure~\ref{fig:thermal_analysis}b) it collapses into a narrow low-intensity peak. This confirms that the spiral arms generally remain in a cool, absorbing (dark fibril) state throughout the observation, consistent with H$\alpha$ fibrils.

 \item \textbf{Threads (bright channels):} The oscillating thread pixels form the brightest population. In the raw distribution (Figure~\ref{fig:thermal_analysis}a), the thread histograms (red/orange/green) are shifted rightward to $> 6.7 \times 10^5$ counts relative to the loops. This separation becomes even more pronounced in the timestep-averaged distribution (Figure~\ref{fig:thermal_analysis}b), where the thread peak is clearly offset from the loop peak with no overlap.

 \item \textbf{Mode dependence:} Both panels suggest a relationship between mode complexity and intensity. Threads with higher-order components extend to the highest intensities, and this effect is clearest in the timestep-averaged distribution (Figure~\ref{fig:thermal_analysis}b), where tertiary-mode threads populate the brightest tail. This supports the idea that regions hosting complex, multi-mode threads are associated with the brightest chromospheric signatures.
\end{enumerate}

These distributions reveal a persistent intensity contrast between the oscillating threads and their surrounding loop structures, consistent with localised energy deposition along oscillatory threads. A quantitative analysis of the associated heating mechanism is deferred to a companion paper currently in preparation. It reveals a sharp, persistent separation: the loops (blue) remain distinctively darker than the background, while the threads (red/orange/green) are consistently the brightest features in the plasma throughout the observation.

\subsection{Curvature-Stratified Oscillation Properties}
\begin{figure*}[t!]
\centering
\includegraphics[width=\textwidth]{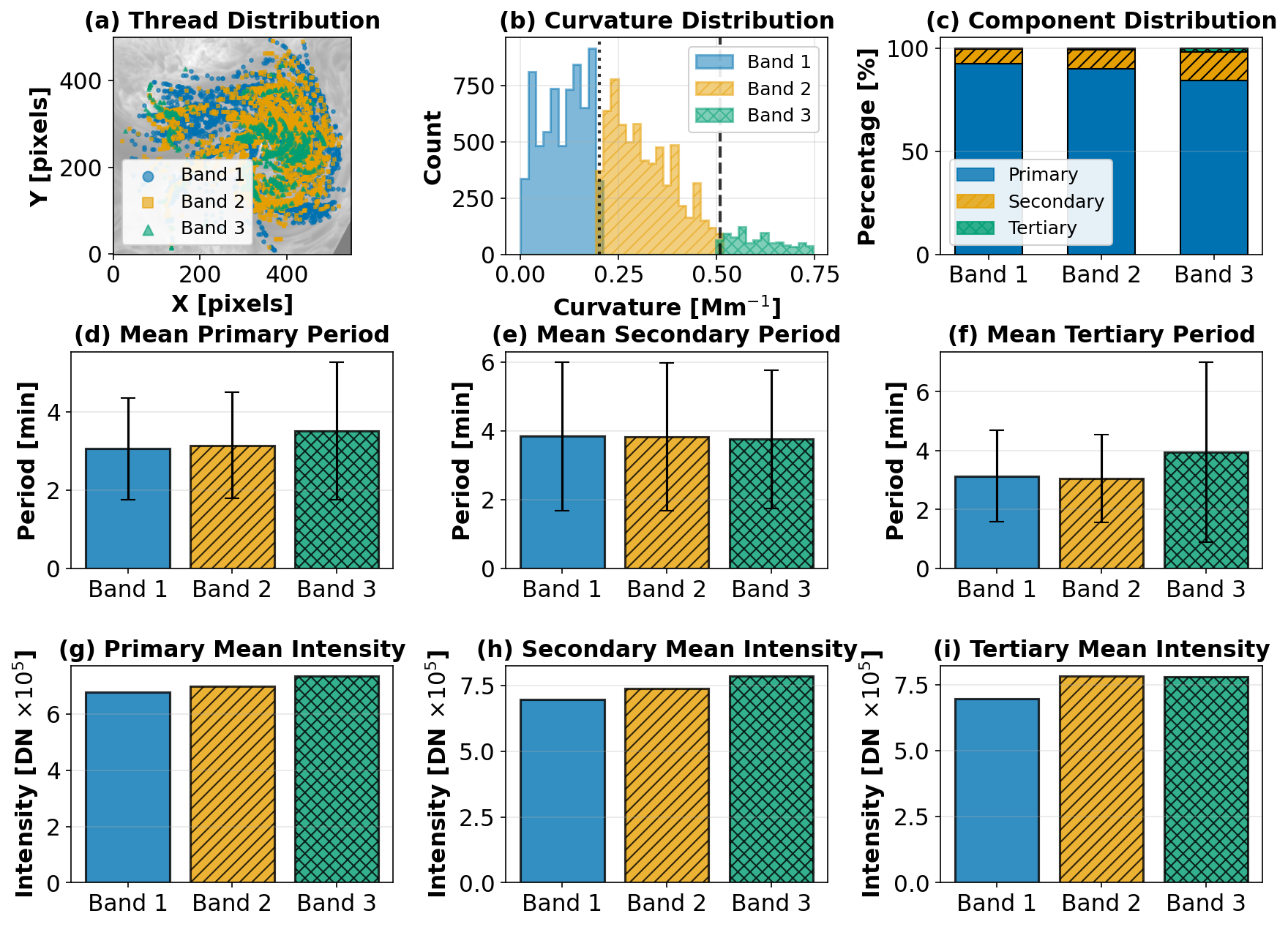}
\caption{Summary of curvature analysis. Threads are grouped into three curvature bands: Band 1 (blue), Band 2 (green), and Band 3 (Red).Row 1: Spatial context and curvature classification. (a) H-alpha chromosphere with spatial distribution of detected threads colour-coded by curvature band. (b) Curvature distribution histogram with threshold boundaries marked by dashed lines. (c) Component distribution of higher order modes relative to curvature bands. Row 2: (d) Mean primary component period in curvature bands. (e) Mean secondary component period in curvature bands. (f) Mean tertiary component period in curvature bands. Row 3: (g) Mean intensity of primary threads in curvature bands. (h) Mean intensity of secondary threads in curvature bands. (i) Mean intensity of tertiary threads in curvature bands.}
\label{fig:curvature_map}
\end{figure*}

To investigate the geometrical influence on the flow dynamics, we classify threads into distinct curvature bands based on the local curvature ($\kappa$) of the pixels the threads occupy. Threads are categorised into three geometric bands: Band~1 (straight, $\kappa \leq 0.2$~Mm$^{-1}$, blue), Band~2 (moderate curvature, $0.2 < \kappa \leq 0.508$~Mm$^{-1}$, green), and Band~3 (high curvature, $\kappa > 0.508$~Mm$^{-1}$, red; Figure~\ref{fig:curvature_map}).

Figure~\ref{fig:curvature_map}a shows how the curvature bands are distributed across the spiral. High-curvature threads (Band~3) are concentrated in and immediately around the pore and the tightest turns of the spiral. Band~2 occupies intermediate geometry and appears both near the pore and along the spiral arms as the structure unwinds. Band~1 dominates the more extended outer sections where the channels are straighter. This separation lets us test whether increased curvature is linked to changes in mode complexity, intensity, and oscillation period.

\subsubsection{Higher-order modes and intensity versus curvature}
Our oscillatory mode-fitting algorithm (Section~\ref{subsec:fitting}) decomposes oscillations into primary, secondary, and tertiary components. We define higher-order modes as threads that exhibit significant secondary or tertiary components. Figure~\ref{fig:curvature_map}c shows that the fraction of higher-order modes increases with curvature, rising from 7.5\% in Band~1 to 10.0\% in Band~2 and 15.7\% in Band~3 (Table~\ref{tab:curvature_statistics}). The mean thread intensity shows the same trend, increasing from 679{,}626 counts (Band~1) to 705{,}325 (Band~2) and 743{,}348 (Band~3). Overall, higher curvature regions support a larger fraction of multi-component behaviour and brighter thread emission signatures (Table~\ref{tab:curvature_statistics}).

\subsubsection{Oscillation period distribution and spatial trend}
Contrary to the standard canopy picture where periods typically increase with distance from the sunspot (i.e. transitioning from 3-minute to 5-minute power as the field expands and rises), we observe the opposite trend in this dataset. In the curvature-stratified period distributions (Figure~\ref{fig:curvature_map}d), the mean primary period increases with curvature, from $3.06 \pm 1.30$~min in Band~1 to $3.15 \pm 1.36$~min in Band~2 and $3.52 \pm 1.76$~min in Band~3 (Table~\ref{tab:curvature_statistics}). Here the quoted uncertainties represent the standard deviation of thread periods within each curvature band. Since the highest-curvature threads are concentrated near the pore (Figure~\ref{fig:curvature_map}a), this corresponds to a preference for longer periods toward the pore ($\sim$3.5~min) and shorter periods in the more extended outer spiral ($\sim$3~min).

\begin{table*}[!h]
\centering
\caption{Curvature Band Statistics}
\label{tab:curvature_statistics}
\small
\begin{tabular}{lccccc}
\hline\hline
Region & Number of  & Mean Curvature & Mean Intensity & Primary Period & Higher-Order \\
       & threads & (Mm$^{-1}$) &  & (min) & Fraction \\
\hline
Band 1  & 6877 & $0.109 \pm 0.057$ & 679626 & $3.06 \pm 1.30$ & 7.5\% \\
Band 2  & 6611 & $0.315 \pm 0.080$ & 705325 & $3.15 \pm 1.36$ & 10.0\% \\
Band 3  & 978 & $0.628 \pm 0.085$ & 743348 & $3.52 \pm 1.76$ & 15.7\% \\
\hline\hline
\end{tabular}
\tablecomments{Higher-order fraction includes secondary and tertiary components.}
\end{table*}

\section{Discussion} \label{sec:discussion}

This study presents the first high-resolution statistical map of flows in a giant chromospheric spiral. By coupling automated thread tracking with oscillatory mode analysis, we have uncovered three primary insights: (1) mode complexity is geometrically influenced by loop curvature; (2) the spiral exhibits an inverse period trend suggestive of a returning loop topology; and (3) the oscillating threads are systematically brighter than their surrounding loop structures, consistent with preferential energy deposition, with a full heating analysis to be presented in a forthcoming companion study.

\subsection{Curvature as a Driver of Thread Complexity}

As noted in our methodology, we apply the OCCULT-2 algorithm \citep{Aschwanden_2013} to trace the spiral flow channels. OCCULT-2 was originally developed and optimised for the relatively smooth, semi-circular geometries of large-scale coronal loops, which legacy TRACE observations demonstrated can reach lengths of $L \approx 100$--$320$~Mm with correspondingly large curvature radii \citep{Aschwanden_2000}. The code enforces a minimum curvature radius threshold ($r_{\rm min} \approx 30$~pixels) to filter out noise and short-lived transients \citep{Aschwanden_2013}, which in the coronal loop context is well-motivated by the known geometry of large-scale EUV loops. Despite this strict filtering, the spiral loops successfully traced in our study geometrically align with modern small-scale loop observations from Solar Orbiter/EUI and SDO/AIA, which have revealed abundant populations of compact loops with lengths of just $\sim$3-30~Mm \citep{Shrivastav2024, Madjarska2024}. The spiral loops exhibit correspondingly extreme 
curvature radii of just $\sim$2-10~Mm (mean curvature values of $\kappa \sim 0.1$--$0.6$~Mm$^{-1}$; Table~\ref{tab:curvature_statistics}), which are an order of magnitude smaller than the large-scale coronal loops for which OCCULT was originally designed. This indicates that the compact, tightly curved geometry of the chromospheric spiral creates a fundamentally distinct oscillatory environment, acting as a unique local catalyst for multi-mode excitation compared to the extended, low-curvature structures traditionally targeted by automated tracing codes.

Beyond the geometrical comparison with coronal loops, we have presented strong observational evidence that magnetic curvature influences oscillatory flow dynamics in the chromosphere. The $2.09\times$ enhancement of higher-order modes in the high-curvature Band~3 compared to the straighter Band~1 suggests that curved magnetic fields are more efficient at exciting or trapping complex oscillatory patterns. As plasma propagates into highly curved regions (Band~3 and Band~2), it may undergo mode conversion or phase mixing, leading 
to the observed higher-order oscillations.

\subsection{Structure of Magnetic Field emerging from Pore}

Our analysis of the spatial distribution of oscillation periods reveals a trend that challenges standard sunspot wave models. Typically, magnetic field lines become more inclined (horizontal) with distance from a sunspot umbra, which raises the acoustic cut-off frequency and results in a transition from 3-minute to 5-minute oscillations \citep{KhomenkoCollados2015_LRSP}. In contrast, we observe longer periods (3--4 min) in the central pore and shorter periods (2--3 min) in the surrounding spiral arms.

\begin{figure*}[t!]
\centering
\begin{subfigure}{0.49\textwidth}
  \centering
  \includegraphics[width=\linewidth]{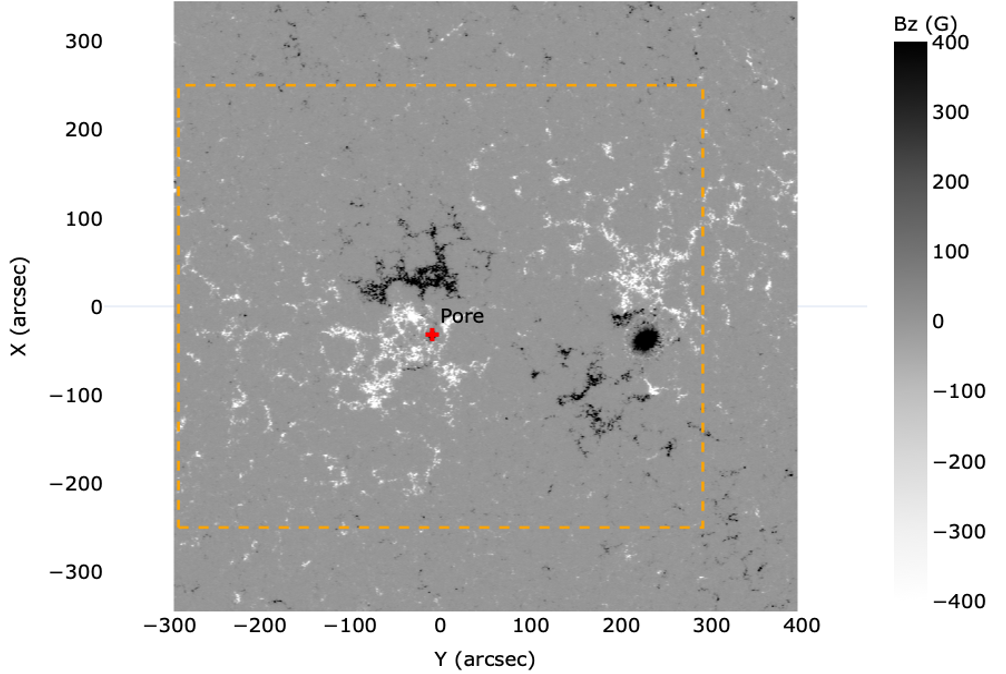}
  \caption{\small HMI extrapolation FOV.}
  \label{fig:mag_a}
\end{subfigure}
\hfill
\begin{subfigure}{0.42\textwidth}
  \centering
  \includegraphics[width=\linewidth]{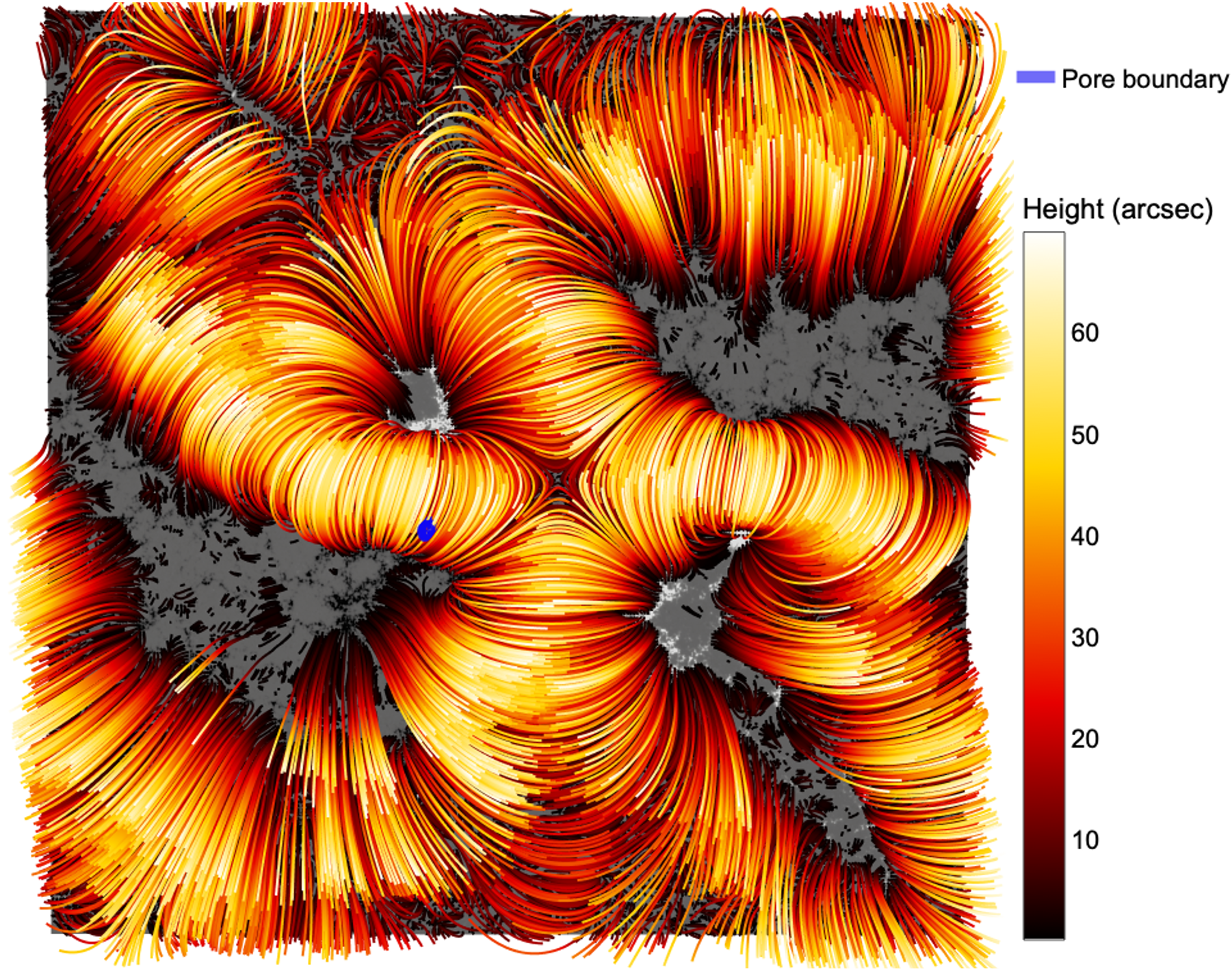}
  \caption{\small Full FOV extrapolation (Max height = 70.0").}
  \label{fig:mag_b}
\end{subfigure}

\vspace{0.5em}

\begin{subfigure}{0.42\textwidth}
  \centering
  \includegraphics[width=\linewidth]{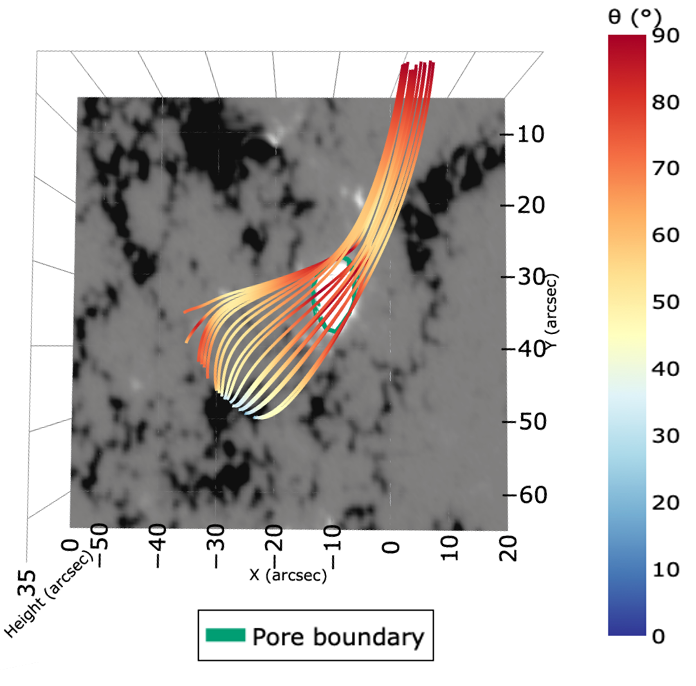}
  \caption{\small Field inclination $\theta$.}
  \label{fig:mag_c}
\end{subfigure}
\hfill
\begin{subfigure}{0.49\textwidth}
  \centering
  \includegraphics[width=\linewidth]{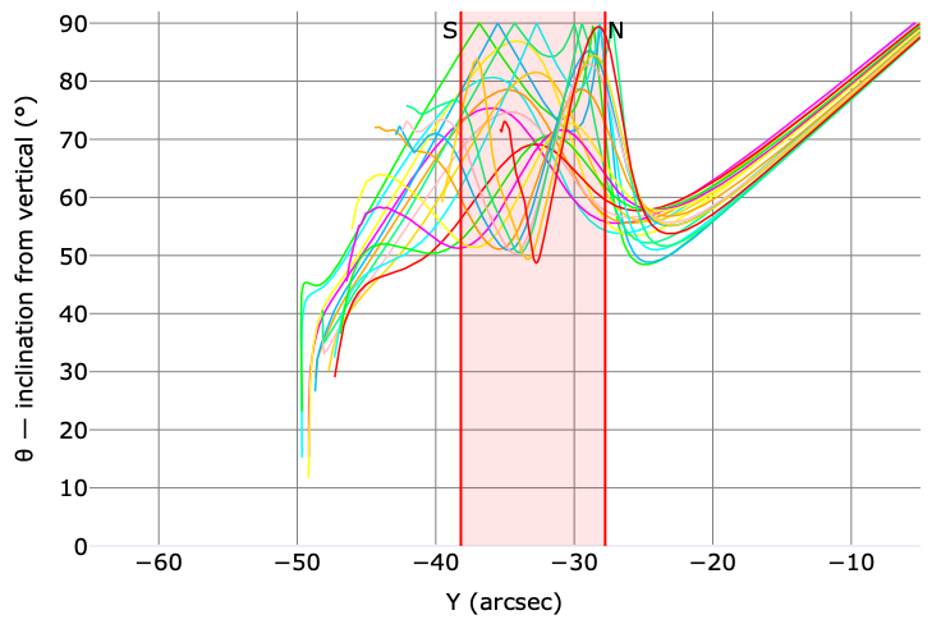}
  \caption{\small Pore-crossing loops (red region is the pore).}
  \label{fig:mag_d}
\end{subfigure}

\caption{Magnetic field structure of the solar pore and its surrounding chromosphere. \textbf{(a)} Full field-of-view HMI line-of-sight magnetogram ($B_z$). The orange box marks the HMI FOV shown in Figure~\ref{fig:sst_observation}, the blue dot is the location of the pore. The red cross indicates the pore centre. \textbf{(b)} Potential field extrapolation performed over the full HMI field of view, with field lines coloured by height above the photosphere. \textbf{(c)} Close-up view of the extrapolated field lines in the vicinity of the pore. Field lines are coloured by inclination angle $\theta$ from the vertical (blue) to horizontal (red). \textbf{(d)} Inclination angle $\theta$ as a function of Y position along pore-crossing field lines. The colours are random and only represent different loops.} \label{fig:magnetic_structure}
\end{figure*}

To explain this anomalous period trend, we must look at the specific magnetic topology of the region. The potential field extrapolation (Figure~\ref{fig:magnetic_structure}; note the magnetogram is rotated $90\degr$ from Figure~\ref{fig:sst_observation}c) demonstrates that the spiral is not a simple expanding funnel, but rather a complex system of closed loops. Crucially, the overlying trans-equatorial quadrupolar structure - previously identified by \citet{Sun_2014} - acts as an overlying magnetic canopy. This canopy compresses the emerging field at chromospheric heights, forcing the field lines directly above the pore into a near-horizontal orientation (mean inclination $\theta = 68.8 \pm 11.1\degr$, Figure~\ref{fig:magnetic_structure}(d)). Because the inclination to the vertical of the magnetic field raises the acoustic cut-off frequency, this canopy effect naturally favours the longer 3-4~minute periods observed above the pore.

Conversely, as these field lines escape the canopy and form the outer spiral arms, they rapidly return to the photosphere. In these returning legs, the magnetic field vector becomes highly radial (vertical) again. This vertical orientation lowers the cut-off frequency, allowing the direct propagation of higher-frequency, short-period (2-3~min) p-modes from the photosphere into the outer spiral arms.

We therefore interpret the observed period gradient as the combined outcome of this unique geometry. The overlying coronal structure shapes the horizontal canopy above the pore, and the emerging flux forms the closed loops that return to the photosphere in the outer spiral. 
Furthermore, the curvature of the spiral structure (Table~\ref{tab:curvature_statistics}) strongly increases toward the pore. Curvature-related effects such as mode conversion and phase mixing provide an additional physical mechanism for modifying the observed oscillatory content across the different regions of the spiral.

The large-scale magnetic context provided by \citet{Sun_2014}, who studied the same trans-equatorial quadrupolar active region using AIA observations and NLFFF extrapolations, supports this interpretation. The emerging flux region they identify in their HMI magnetograms is spatially consistent with the magnetic pore studied here, suggesting a possible multi-scale connection in which the same flux emergence process drives both the chromospheric spiral at lower heights and the large-scale coronal activity identified by \citet{Sun_2014}. Their finding that the NLFFF structure closely resembles the potential field in the high corona is also broadly consistent with our own extrapolation (Figure~\ref{fig:magnetic_structure}), lending further confidence to the canopy geometry and period gradient interpretation presented above.

\section{Acknowledgments}
All authors acknowledge the UK Science and Technology Facilities Council (STFC) for IDL support. Y.S. would like to thank Northumbria University for the award of PhD studentship funding through the internal RDF Programme. 

The research was sponsored by the DynaSun project and has thus received funding under the Horizon Europe programme of the European Union under grant agreement (no. 101131534). Views and opinions expressed are however those of the author(s) only and do not necessarily reflect those of the European Union and therefore the European Union cannot be held responsible for them. This work was also supported by the Engineering and Physical Sciences Research Council (EP/Y037464/1) under the Horizon Europe Guarantee.

\vspace{5mm}
\noindent
\textit{Facilities:} The Swedish 1-m Solar Telescope (SST) is operated on the island of La Palma by the Institute for Solar Physics of Stockholm University in the Spanish Observatorio del Roque de los Muchachos of the Instituto de Astrof\'{ı}sica de Canarias.

\vspace{5mm}
\noindent
E.S. would like to pay special thanks to Pit S\"{u}tterlin for his support during the observation campaign. 

\vspace{5mm}
\noindent
\textit{Software:} CRISPEX \citep{Vissers_2012}

\clearpage
\bibliography{ref_review_2}
\bibliographystyle{aasjournal}

\end{document}